\documentstyle[12pt]{article}
\begin{document}
\title{Stability of Compacton Solutions of Fifth-Order Nonlinear  
Dispersive Equations} 
\author{Bishwajyoti Dey \\ Department of Physics \\
University of Pune \\ Pune - 411 007 , India \\and\\ Avinash Khare\\
Institute Of Physics\\ Bhubaneswar - 751005 ,India.}
\maketitle
\begin{abstract}
We consider fifth-order nonlinear dispersive $K(m,n,p)$ type equations 
to study the effect of nonlinear dispersion. Using simple scaling arguments 
we show, how, instead of the conventional solitary waves like solitons, the 
interaction of the nonlinear dispersion with nonlinear convection generates 
compactons - the compact solitary waves free of exponential tails. This 
interaction also generates many other solitary wave structures like cuspons, 
peakons, tipons etc. which are otherwise unattainable with linear dispersion. 
Various self similar solutions of these higher order nonlinear dispersive 
equations are also obtained using similarity transformations. Further, it is 
shown that, like the third-order nonlinear $K(m,n)$ equations, the 
fifth-order 
nonlinear dispersive equations also have 
the same four conserved quantities and further even any 
arbitrary odd order nonlinear dispersive $K(m,n,p...)$ type equations also 
have the same three (and most likely the  
four) conserved quantities. Finally, the stability of the compacton 
solutions for the fifth-order nonlinear dispersive equations are 
studied using linear stability analysis.
From the results of the 
linear stability analysis it follows that, unlike solitons, all the allowed 
compacton solutions are stable, since the stability conditions are satisfied 
for arbitrary values of the nonlinear parameters.  

\end{abstract}
\newpage

\section{Introduction}

The recent discovery [1] that solitary wave solutions supported by nonlinear
wave equations may compactify under nonlinear dispersion, has shown 
that nonlinear dispersion can cause qualitative changes in the nature 
of genuinely nonlinear phenomena. Such nonlinearly dispersive 
partial differential equations which support compacton solutions
are represented by the $K(m,n)$ equations of the form
\begin{equation}
u_t + a(u^m)_x + (u^n)_{3x} =0; \hspace{.2in}m,n>1
\end{equation}

Most of the 
weakly nonlinear and linear dispersion equations studied so far 
admit solitary waves, called solitons, that are infinite in extent.
On the other hand, it has been 
shown that the interaction of nonlinear dispersion with nonlinear 
convection generates exactly compact structures, called compactons, 
free of exponential tails. The compacton solutions so generated 
have immediate applications in the study 
of pattern formations, as the observed stationary and dynamical patterns 
in nature are usually finite in extent. The interaction also generates 
many other nonlinear solitary wave structures like cuspons, peakons, 
tipons etc. [2] which are otherwise not possible in the weakly nonlinear 
models with linear dispersion. The compacton speed 
depends on its height, but unlike solitons, its width is independent of 
its speed. Beside the compact structure and the unusual speed-width 
relation, the compactons have the remarkable soliton like property that they 
collide elastically. However, unlike soliton collisions in an 
integrable system, the point at which the compactons collide is marked 
by the creation of low amplitude compacton-anticompacton pairs.More 
recently [3,4] the study of the third-order $K(m,n)$ equations was 
generalised by including into the equation higher order nonlinear 
dispersive terms, like for example, the fifth-order $K(m,n,p)$ equations 
of the form [3,4]
\begin{equation}
u_t + \beta  _1 {(u^m)}_x + \beta _2 {( u^n)}_{3x} +
\beta _3 {(u^p)}_{5x} =0, \hspace{.2in}m, n, p >1
\end{equation}

This type of higher order dispersive 
equations are useful for describing the dynamics of various physical 
systems. As has been explained in [1], the compacton supporting 
nonlinear dispersive equations arises if one controls the effects of 
nonlinearity and 
dispersion by two independent parameters while modelling physical 
phenomena. In such cases, it is required to retain the quadratic and higher 
order effects in dispersion. For example, one can show that the equations 
governing the motion of mass points in a dense chain (which 
leads to vibrational excitations in the chain) are a prototype of 
the Eq.(2) above, if one consider the effect of the higher 
order dispersive interactions. Similarly, for plasma ion acoustic waves, 
if the ion-electron charge separation is treated separately from the 
ions inertia, then again, considerations of 
the effects of the higher order dispersive 
interactions leads to the equations of the form as in Eq.(2) above.   
Beside the applications of such higher order equations in physical systems (see
 [3,4] for details), the studies of these generalised equations such as the 
fifth-order nonlinear dispersive equations as above (Eq.(2)) are motivated by 
the need to understand how far the concept of compact structures can be 
extended and how generic are the properties of the compacton solutions. 
A variety of explicit compact solitary wave structures of these fifth- 
order nonlinear dispersive equations are constructed [3,4] and numerical 
simulations of these equations have also revealed the existence of 
compact travelling breathers [4]. 

In the present paper we report on the study of the stability property of the 
compacton solutions [3,4] of these fifth-order nonlinear 
dispersive equation.Beside predicting about the asymptotic nature of the 
compacton solutions, the stability analysis is also important for the following
 reasons.
Since the nonlinear dispersive equations represented by $K(m,n)$ 
equations (Eq.(1))
 do not appear to be integrable [1], this suggests 
that the observed almost elastic collisions of the compactons are 
probably not due to the integrability property and thus the mechanism 
responsible for the compact structure, coherence and robustness of the 
compactons 
calls for a more systematic study of the nonlinear 
dispersive systems.
Stability analysis of the compacton solutions may 
provide some clues regarding 
 the almost elastic nature of the compacton collisions. Beside, the stability 
problem of the nonlinear dispersive equation is interesting 
because for such equations 
with higher power of the nonlinearity and nonlinear dispersion, the phenomena of 
collapse is possible. 
Further motivation for studying the stability properties of the 
fifth-order nonlinear dispersive equations comes from the result of 
the recent study on the 
role of the fifth-order dispersion term on the soliton stability of 
the usual KdV type linear dispersion equations. For example, it has 
been shown that [5] the solitary wave solutions of the 
fifth-order equation of the type
\begin{equation}
u_t + u^pu_x +\alpha u_{3x} + \beta u_{5x} =0
\end{equation}
are unstable with respect to the collapse 
type instabilities, if p $\geq$ 4 for $\beta $ =0 while for 
$\beta \neq $ 0, i.e. the addition of the fifth-order term stabilizes 
the soliton for p$>$4 [5].
 The exact upper limit of the nonlinearity  
parameter $p$ in this case is still an open question [6]. 
It would therefore be appropriate to examine whether the addition of the 
higher 
order nonliner dispersion term puts any additional constraint on the conditions
for the stability of the corresponding compacton solutions. 
Recently some attempts have been made to numerically 
study the stability of compacton solutions of the fifth-order nonlinear 
dispersive equations [7]. 

The rest of the paper is organised as follows: in Sec. 2  we present 
some general properties like various self similar solutions, 
conservation laws, various solitary wave structures etc. 
of the fifth-order nonlinear dispersive 
equations.
In Sec. 3  we discuss 
the stability of the compacton solutions of these 
equations using linear stability analysis. A short report on this 
 method appeared recently 
in an rapid communication article [8].
Finally we conclude in Sec. 4. 

\section{Some General Properties}

The fifth-order nonlinear dispersive $K(m,n,p)$ equation of the 
form as in Eq.(2)
 is not 
derivable from a Lagrangian and hence does not 
possess the usual conservation laws of mass, energy etc. that are 
associated with the KdV type of equations ($m=n=p=1$ case). However, this 
equation has exact compacton solutions for the nonlinearity parameters 
within the range 2$\leq$ k $\leq$ 5, provided $k=m=n=p$ [3,4]. Since 
this equation 
does not have a Lagrangian and hence a conserved Hamiltonian, we cannot do 
a linear stability analysis for the compacton solutions of this equation,
as the linear stability requires a Hamiltonian, as shown below. Hence, 
we consider a slightly different fifth-order nonlinear dispersive equation 
\begin{eqnarray}
\lefteqn{u_t  - a \delta u^{a - 1} u_x + \alpha b (b-1) u^{b-2} (u_x)^3 +
4 \alpha b u^{b-1} u_x u_{2x}  + 
 2 \alpha u^b u_{3x} } \nonumber \hspace{-.1in}\\
&  & \mbox{} + 3 \beta c (c -1) u^{c-2} (u_x)^5 + 24 \beta c u^{c-1} (u_x)^3 u_{2x}
+ 24 \beta u^c u_x (u_{2x})^2  \nonumber \\
&  &  \mbox{} + 12 \beta u^c (u_x)^2 u_{3x} - 
2 \gamma d (d-1)(d-2) u^{d-3} (u_x)^3 u_{2x} \nonumber \\
&  & \mbox{} -
7 \gamma d (d-1) u^{d-2} u_x (u_{2x})^2  
 - 6 \gamma d (d-1) u^{d-2} (u_x)^3 u_{3x} - 
10 \gamma d u^{d-1} u_{2x} u_{3x}  \nonumber \\
& & \mbox{} - 6 \gamma d u^{ d-1} u_x u_{4x} - 2 \gamma u^d u_{5x} = 0
\end{eqnarray}
where $u(x)=\partial _x \phi (x)$. 
At a first glance it may appear that this equation has been created 
artificially.
However, a closer look will immediately reveal that this equation is very 
similar to Eq.(2) above. It contains exactly the same terms as in Eq.(2), 
but only the weightage of the terms are different. Comparing term by term 
we see that  
the set of parameters ${m,n,p}$ in Eq. (2) 
corresponds to $a=m, b+1=n$ and $c+3=d+1=p$ in Eq. (4). Both Eqs.(2) and (4) 
have compacton solutions. But the advantage of Eq.(4) is that it is 
possible now to write a Lagrangian for this equation leading to a Hamiltonian 
and thus one can do the stability analysis of the compacton solutions of 
Eq.(4). Thus we will present here the stability analysis of the 
fifth-order nonlinear dispersive equation given by Eq.(4), instead of 
Eq.(2).  The Lagrangian corresponding to Eq.(4) is given by 
\begin{eqnarray}
\lefteqn{\hspace{-.25in} L  =  \int {\cal L}\, dx }\nonumber  \\
&  &  \hspace{-.5in} = 
\int dx \big{[} \frac{1}{2} \phi _x \phi _t - 
\delta \frac{{(\phi _x)}^{a +1}}{a+1} - \alpha {(\phi _x)}^b 
{(\phi _{2x})}^2 - \beta {(\phi _x)}^c{(\phi _{2x} )}^4 - 
\gamma {(\phi _x)}^d {(\phi _{3x})}^2 \big{]}
\end{eqnarray}
It is worth remarking here that recently Cooper et al [7] 
have obtained compacton
solutions for a
slightly different fifth-order nonlinear equation which is again obtainable
from a Lagrangian. 

The conserved Hamiltonian
H which is obtained from the Lagrangian [Eq. (5)] is
\begin{eqnarray}
\lefteqn{\hspace{-.5in} H= \int_{-\infty }^{\infty } 
[ \pi \dot {\phi}  - {\cal L}] dx, \hspace{.5in} \pi = 
\frac{\partial {\cal L}}
{\partial {\dot \phi}} = \frac{1}{2} \phi _x} \nonumber \\
& & \hspace{-.6in}=\int _{-\infty }^{\infty} 
\bigg [ \delta \frac{ u^{a+1}}{a+1} 
+ \alpha u^b (u_x)^2 + \beta u^c
(u_x)^4 + \gamma u^d (u_{2x})^2 \bigg ] dx
\end{eqnarray}
This Hamiltonian also follows from the fact that Eq. (4) can be
written in the canonical form
\begin{equation}
u_t = \partial _x \frac{ \delta H}{\delta u} =\{u, H\}
\end{equation}
where the Poisson bracket structure is given by
\begin{equation}
\{ u(x) , u(y) \} = \partial _x \delta (x -y)
\end{equation}

We first consider the scaling relations between speed, width and amplitude 
of the travelling wave $(\xi =x+Dt)$ structure of Eq. (4). Under the 
scaling transformation $x \rightarrow \mu x, t \rightarrow \nu t$ and 
$u \rightarrow \eta u$, it can be easily shown that
\begin{equation}
u(x,t) \rightarrow D^{\frac {1}{a-1}}u(D^{\frac {(b+1-a)}{2(a-1)}} \xi ) = 
D^{\frac {1}{a-1}} u (D^{\frac {(c+3-b-1)}{2(a-1)}} \xi ) =   
D^{\frac {1}{a-1}} u (D^{\frac {(d+1-b-1)}{2(a-1)}} \xi )
\end{equation}
Similarly Eq. (2) under scaling transformation also admits solutions of the 
form
\begin{equation}
u(x,t) = D^{\frac {1}{(m-1)}} u ( D^{\frac {(p-n)}{2(m-1)}} \xi) 
\end{equation}
From Eq. (9) we see that for the case when $a=b+1=c+3=d+1$ (which 
corresponds to the case of $m=n=p$ in Eq. (10)), there is a 
detailed balance between the convection and dispersion and as a 
result of which the width of the sotitary wave solutions (compactons) 
become independent of the amplitude (or speed $D$).  

It can also be shown that the fifth-order equations [Eq. (4)] are invariant 
under the stretching group of transformations [2] which supports 
self-similar solutions (similarity structures) of the form
\begin{equation}
u(x,t) = t^{\frac {2}{\Delta}} F(\zeta )
\end{equation}
where $\zeta =xt^{-\mu},\  \mu ={\frac {b+1-a}{\Delta}}$ and $ \Delta =
b+3(1-a)$ along with $c+3=d+1$. Now, when $\mu =0$ or $\Delta =0$, the 
self similar soluions of the form in Eq. (11) is no longer valid. 
However, in that case, it can be shown that the equations [Eqs. (2) and (4)] 
have additional invariance under shifts in time or space (spiral symmetry) 
giving rise to the self similar solutions of the form
\begin{eqnarray}
u(x,t) & = & t^{\frac {1}{1-a}}F(x+Dlogt)\hspace{.2in}\  for \ a=b+1=c+3=d+1 \nonumber \\ 
       & = & e^{-Dt}F(xe^{D(a-1)t}) \hspace{.2in} \ for \ 3a=b+3
\end{eqnarray}
Similarly, it can be shown that the fifth-order nonlinear dispersive $K(m,n,p)$ 
equations 
of the form of Eq. (2) are also invariant under stretching group of 
transformations 
which admits similarity solutions of the form as in Eq. (11) 
for $\mu = (p-n)/\Delta $ where 
$\Delta = 3p-5n+2$ and $m=2n-p$. Again, when $n=m$ or when 
$n=3m-2$ it can be shown that the $K(m,n,p)$ equations are invariant under 
the spiral group of transformations leading to self similar solutions 
of the form
\begin{eqnarray}
u(x,t) & = & t^{\frac {1}{1-m}}F(x+Dlogt) \hspace{.2in}\ for \ m=n=p \nonumber \\
       & = & e^{-Dt}F(xe^{D(m-1)t}) \hspace{.3in} \ for \ 5n=3p+2 \ and \ m=2n-p
\end{eqnarray}
For the case when $m=n+2$ and $p=n-2$, substituting Eq. (11) in Eq. (2) we 
get the equation for $F(\xi)$ as
\begin{equation}
-\alpha \xi F + F^{n+2} + (F^n)_{2\xi } + (F^{n-2})_{4\xi } = \beta
\end{equation}
where $\xi = xt^{-\alpha }$ , $\alpha = {\frac {1}{n+2}}$ and $\beta $ =constant.
For the particular case of $n$=2 ( in which case Eq. (2) reduces to the form 
of the third- 
order nonlinear dispersion $K(m,n)$ equation [Eq. (1)]) Eq. (14) reduces 
to the generalised second-order Painleve equation [2]. 

We now make some comment
 about the conserved quantities associated with these fifth-order nonlinear 
dispersive equations.
 A conservation law associated with equations of the form Eq.(2) and Eq.(4) 
can be written as
\begin{equation}
\frac {\partial Q}{\partial t} + \frac {\partial X}{\partial x} =0
\end{equation}
where $Q$ is the density of the conserved quantity $\int_{-\infty}^{\infty}Q dx$
 and $X$ is the corresponding flux density.
In an earlier paper [3] we had reported 
that Eq. (2) has only one conservation law. We now find that, like the $K(m,n)$ 
equations (Eq. (1)), 
the fifth-order $K(m,n,p)$ 
equations (Eq. (2)) also have four conservation laws for the case   
$m=n=p$, with four densities (Q's) same as that of Eq.(1) [1].  
In fact, we are now able to show  
that the same three (out of four) densities (Q's) of Eq.(1) 
also exist 
for arbitrary odd ($2n+1$)th-order nonlinear dispersion 
equations (Eq. (36) in [3]).
It can be checked that 
the $Q$ and $X$ values for the arbitrary odd ($2n+1$)th- 
order $K(m,m,m,...)$ 
equations are 
given by
\begin{eqnarray} 
\noindent
Q_1 =  u \ , \ \hspace{.3in}
X_1 & = & a_1(u^m) 
+ (a_1 +a_3)(u^m)_{2x} + .... \nonumber \\ 
    & + & (a_{2n-3} + a_{2n-1})(u^m)_{(2n-2)x} 
    + a_{(2n-1}(u^m)_{(2n)x}      \nonumber \\
Q_2 =  ucosx \ , \ \hspace{.1in}
X_2 & = & a_1 [\sin x (u^m)_x + \cos x (u^m)_{2x}] +... \nonumber \\ 
    & + & a_{2n-1}[\sin x (u^m)_{(2n-1)x}  
    + \cos x (u^m)_{(2n)x}]      \nonumber \\
Q_3 =  u sinx  \ , \ \hspace{.1in}
X_3 & = & a_1[-\cos x (u^m)_x + \sin x (u^m)_{2x}] +... \nonumber \\ 
    & + & a_{(2n-1)}[-\cos x (u^m)_{(2n-1)x}  
+ \sin x (u^m)_{(2n)x}]      
\end{eqnarray}
where $a_1$, $a_3$ etc are coefficients of various dispersive terms.
We have not been able to prove in general 
the existence of the fourth density
$Q_4=u^{m+1}$ [1] for the arbitrary odd order nonlinear equation 
. However, we can show that for the third and fifth-order 
nonlinear dispersion equation
 $Q_4 = u^{m+1}$ is the fourth density
for arbitrary values of the nonlinearity parameter $m$. 
 We have also checked that $Q_4=u^{m+1}$ is also the fourth 
density for the ninth and eleventh-order equation for the 
particular value of the nonlinearity parameter $m=2$. 
 From the above 
calculations we conjecture that $Q=u^{m+1}$ is the fourth density
for the arbitrary odd order nonlinear dispersion equation 
for arbitrary values of the nonlinearity parameter $m$.
The origin or the symmetry associated with these unusal  
conservation laws of these nonlinear dispersive equations are not known 
at present. However, it should be noted that for the fifth-order nonlinear 
dispersive 
equation [Eq. (4)] which is derivable from a Lagrangian, there are only 
three conserved quantities.

It can be easily shown [3,4] that both the equations Eq.(2) and Eq.(4)  
support a class of one parameter 
family of compacton solutions of the form 
\begin{equation}
u(\xi ) = A \cos^{\nu}(B\xi )
\end{equation}
for$ \mid B\xi \mid \leq \pi /2 $, $u(\xi ) =0$ otherwise and where 
$\xi = x- Dt \ , \  \nu = \frac {4}{(k-1)}.$ The width B of the 
compacton solutions is independent of speed D. The compacton solutions 
exist for the continuous values of the nonlinearity parameter $k=a=
b+1=c+3=d+1$ in the range $2\leq k \leq 5$. Similarly, it can be shown 
that the fifth-order nonlinear dispersive equations of the form of 
$ K(m,n,p)$ equations (Eq. (2)) also have the compacton solutions of the form 
of Eq. (17) within the same range of the nonlinearity parameter 
$2 \leq k=m=n=p \leq 5$. For the third-order nonlinear dispersive 
equations (Eq.(1)) it has been shown that, 
 beside compacton solutions, the interaction of the  nonlinear dispersion term 
with the convective term gives rise to various other kinds of nonlinear 
localised structures such as cuspons, peakons, spikons, tipons etc. [2]. 
 In the same way, we have been able to obtain the peakon solution of the form
\begin{equation}
u(x,t) = u_0 [e^{-\beta {\mid \xi \mid} } -1]
\end{equation}
for the fifth-order nonlinear dispersive equation (Eq.(2)), but, only for the 
special case of the nonlinear parameters $m = n = p = 2$. 
So far we have not been able 
to obtain the peakon solutions (if any) for the same equations for other 
values of the nonlinear parameters $m,n,p$. 
Similarly, we have not yet been able to 
obtain other solitary wave structures (if any) 
such as cuspons and tipons for the 
fifth-order nonlinear dispersive equations (Eqs. (2),(4)).

\section{Compacton Stability}
In this section we discuss in detail the stability analysis of the compacton 
solutions of the fifth-order nonlinear dispersive equations which 
have been studied recently [3,4]. We use 
here the method of linear stability analysis to analyze the problem. 
As 
has been mentioned above,  we cannot do the linear stability analysis 
for the compacton solutions of Eq.(2) as these equations do not have a 
Lagrangian. Accordingly we will study the stability of the compacton 
solutions for a slightly different equation (Eq.(4)) which can be 
derived from a Lagrangian density 
(Eq. (5)) and has a conserved Hamiltonian [Eq. (6)] and 
momentum 
\begin{equation}
P(u) = \frac {1}{2} \int_{-\infty}^{+\infty} u^2 dx
\end{equation}
It can be easily checked that Eq. (4) can also be derived from the variational 
principle $\delta (H+Dp) =0$, where $P$  and $D$ denote the compacton 
momentum and velocity respectively. Introducing the notations
\begin{eqnarray}
I_n & = & \int_{-\infty}^{+\infty} u^n (x) dx, 
\ \ J_2 = \int_{-\infty}^{+\infty} 
u^b (u_x)^2 dx, \ \ J_3 = \int_{-\infty}^{+\infty} u^c (u_x)^4  dx \nonumber \\ 
J_4 & = & \int_{-\infty}^{+\infty} u^d (u_{2x })^2 dx
\end{eqnarray}
we can write the Hamiltonian [Eq. (6)] and the momentum for the compactons as
\begin{equation}
H_c = [ \frac {\delta}{(a+1)} I_{a+1} +\alpha J_2 + 
\beta J_3 + \gamma J_4] \, , \ \ P_c =\frac {1}{2} I_2
\end{equation}
Now we consider the scaling transformation $x\rightarrow \nu x$. Under this 
transformation the integrals in Eq. (20) are transformed as 
\begin{equation}
I_n(\nu )= \frac {I_n}{\nu }, \ \ J_2(\nu )= \nu J_2, \ \ J_3(\nu ) = 
\nu ^3 J_3 \ \ and \ \ J_4(\nu ) = \nu ^3 J_4
\end{equation}
such that 
\begin{equation}
H_c(\nu ) = \frac {\delta }{\nu (a+1)} I_{a+1} + \alpha \nu J_2 + \beta \nu ^3 
J_3 + \gamma \nu ^3 J_4, \ \ and \ \ P_c(\nu )= \frac {P_c}{\nu }
\end{equation}
Integrating Eq. (4) twice we get
\begin{equation}
2DP_c + \delta I_{a+1} +\alpha (b+2) J_2 + \beta (c+4) J_3 +\gamma (d+2) J_4 =0
\end{equation}
Similarly from
$$\frac {d}{d\nu } [ H(\nu ) + D P_c (\nu )]_{\nu =1} =0 $$  
we get,
\begin{equation}
 -\frac {\delta }{(a+1)} I_{a+1} + \alpha J_2 + 3\beta J_3 + 3\gamma J_4 
-DP_c=0   
\end{equation}
We eliminate $J_4$ and $I_{a+1}$ from Eqs. (24) and (25) to 
write the Hamiltonian (Eq. (21)) as 
\begin{eqnarray}
H_c & = & \frac {\alpha J_2}{(3a+d+5)} (2a-4b+2d-2)+\frac {J_3}{(3a+d+5)} 
(-8\beta 
+4\beta d -4c\beta ) \nonumber \\ 
   & = & \frac {DP_c}{(3a+d+5)}(a-d-9) \nonumber
\end{eqnarray}
However, as mentioned above, the compacton solutions (Eq. (17)) exist for 
nonlinear parameters $a=b+1=c+3=d+1=k$ within the range $2\leq k \leq 5$  for 
which the Hamiltonian can be written as
\begin{equation}
H_c = - \frac {2DP_c}{(a+1)}
\end{equation}
Similarly Eq. (23) can be written as
\begin{equation}
H_c(\nu )=\frac {\alpha J_2}{2}(2\nu -1/\nu -\nu ^3) + \frac {DP_c}{4(a+1)} 
[\nu ^3 (a-1) - \frac {(a+7)}{\nu }]
\end{equation}
Thus $H_c(1)=H_c$. Now, we consider the more general scaling transformation 
$u \rightarrow \mu ^{\frac {1}{2}}u(\lambda x)$. Under this transformation 
$H_c$ and $P_c$ are transformed as $H(\lambda ,\mu )$ and $P(\lambda , \mu)$ 
and 
\begin{eqnarray}
\Phi (\lambda , \mu ) & = & \frac {\delta }{\lambda (a+1)} 
\mu ^{\frac {a+1}{2}}I_{a+1} 
+\alpha \lambda \mu ^ {\frac {b+2}{2}}J_2 + \beta \lambda ^3 \mu ^ {\frac {c+
4}{2}}J_3 \nonumber \\ 
                      &   & \mbox{} + \gamma \lambda ^3 \mu ^{\frac {d+2}{2}}J_4 + D\frac {\mu }{\lambda }P_c
\end{eqnarray}
where $\Phi (\lambda , \mu)  = H_c(\lambda ,\mu )+DP_c (\lambda , \mu )$. The 
expressions $\frac {\partial \Phi}{\partial \lambda }= \frac{\partial \phi }{
\partial \mu }=0 $ give the stationary points at $\mu =\lambda =1$ (the 
compacton equation) and near this point, using the Taylor's series for $\mu = 
\lambda $ we get (the transformation in this case does not change the 
momentum $P$)
\begin{eqnarray}
\delta ^{(2)}\Phi (\lambda ) & = & \delta ^ {(2)} H(\lambda ) = 
\frac {(\lambda -1)^2}{8} 
\bigg [\frac {\delta I_{a+1}}{a+1}(a-1)(a-3) \nonumber \\ 
                             & + & \alpha J_2 
(b+2)(b+4) \mbox{} 
+ \beta J_3 (c+10)(c+8)  \nonumber  \\
                             & + & \gamma J_4 (d+6)(d+8) \bigg ]
\end{eqnarray}
which has a definite sign. If it is positive (negative), the expression
\begin{equation}
H_c(\lambda )=H_c(\mu ,\lambda )|_{\mu  = \lambda }=\frac {\delta }{a+1} 
\lambda ^{\frac {(a-1)}{2}}I_{a+1} +\alpha \lambda ^{\frac {(b+4)}{2}}J_2 + \beta 
\lambda ^{\frac {(c+10)}{2}}J_3 +\gamma \lambda ^{\frac {(d+8)}{2}}J_4
\end{equation}
has a minimum (maximum) at $\lambda =1$.

Now, let us assume that $u=u_c + v$, where $\mid v \mid \ll 1$ and the scalar 
product $<u_c,v>=0$. Substituting this in Eq. (4) and after linearization we 
get
\begin{equation}
\partial_T v=\partial _{\xi } \hat {L}v
\end{equation}
where $\xi =x-Dt \ , \ T=t$ and the operator $\hat {L}$ is given by
\begin{eqnarray}
\hat {L} & = & D + a\delta u^{a-1}-2\alpha bu^{b-1}u_x\partial _x -\alpha b(b-1)
u^{b-2}u_x^2 -2\alpha bu^{b-1}u_{2x} -2\alpha u^b \partial _x ^2 \nonumber \\
         & - & 3\beta c(c-1)
u^{c-2}u_x^4 -12\beta cu^{c-1}u_x^3\partial _x -12 \beta cu^{c-1}u_x^2u_{2x} 
 -24\beta u^cu_xu_{2x}\partial _x \nonumber \\ 
         & - & 12\beta u^cu_x^2\partial _x^2 
+ 3\gamma d(d-1)u^{d-2}u_{2x}^2 +6\gamma du^{d-1}u_{2x}\partial _x^2  
+ 2\gamma d
(d-1)(d-2)u^{d-3}u_x^2u_{2x} \nonumber \\ 
         & + & 4\gamma d(d-1)u^{d-2}u_xu_{2x}\partial _x  
+2\gamma d(d-1)u^{d-2}u_x^2\partial _x^2 
 +4\gamma d(d-1)u^{d-2}u_xu_{3x} \nonumber \\ 
         & + & 
4\gamma d u^{d-1}u_{3x}\partial _x 
          +4\gamma du^{d-1}u_x\partial _x^3 +
2\gamma du^{d-1}u_{4x} +2\gamma u^d \partial _x^4
\end{eqnarray}
Eq. (31) has a solution of the form
\begin{equation}
v(\xi ,T) = e^{-iwT}\phi (\xi ) + e^{iw^*T}\phi ^*(\xi )
\end{equation}
where $ \phi (\xi )$ satisfies the equation 
\begin{equation}
w\phi (\xi ) = i\partial _{\xi } \hat{L} \phi (\xi )
\end{equation}
Integrating the compacton equation of motion (Eq. (4)) once w.r.t. $\xi =
x-Dt$, the resulting equation can be written as 
\begin{equation}
\hat{L}\partial _{\xi }u_c =0
\end{equation}
Similarly, integrating Eq. (4) once w.r.t. $\xi $ and differentiating 
the resulting equation w.r.t $D$ we get
\begin{equation}
\hat{L}(\frac {\partial u_c}{\partial D})=-u_c
\end{equation}
Eq. (35) shows that the $w=0$ 
solution of Eq. (34) is given by $\phi (x) \propto 
\partial _{\xi } u_c$. Similarly, Eq. (34) has also the solution 
$(-w, \phi (-\xi ))$. Thus the compacton $u_c$ is stable if $w$ is real and 
unstable if $w$ is complex.

Since Eq. (34) contains the product of two hermitian operators, hence all $w$ 
are real if one of the operators is positive definite. This implies that 
a sufficient condition condition of real eigenvalue $w$ is [5]
\begin{equation}
<\psi , \hat{L} \psi > \ \ >0
\end{equation}
where $\psi $ is a function in the the subspace orthogonal to $u_c$, i.e.
\begin{equation}
<\psi , u_c > = <\psi , \partial _{\xi }u_c >=0
\end{equation}
Using Eqs.(35) and (36) and following Karpman [5], it can be shown that 
the condition for the existence of such function $\psi $ satisfying Eqs. (37) 
and (38) is equivalent to the condition
\begin{equation}
(\frac {\partial P_c}{\partial D}) >0
\end{equation}
From Eqs. (20), (21) and the exact compacton solution [Eq. (17)] of the 
fifth-order nonlinear dispersive equation [Eq. (4)], it can be shown that 
the condition in Eq. (39) is satisfied for arbitrary values of the nonlinear 
parameters $k=a=b+1=c+3=d+1$. However, as has 
been mentioned above, the compacton 
solutions of the fifth-order nonlinear dispersive equation are allowed for 
the nonlinearity parameter in the range $2\leq k \leq 5$. Since the stability 
condition [Eq. (39)] is satisfied for arbitrary values of the nonlinearity 
parameter $k$, this implies that all the allowed compacton solutions [Eq. (17)] 
are stable.

We can obtain another condition for the compacton stability from the 
Hamiltonian minimum condition. This is because the condition in Eq. (37) 
is also associated with the extremum of $H+DP$, since, using the relation 
$\delta (H+DP)=0$, one can show that the second variation of $H(u)$ and 
$P(u)$ at $u=u_c$ can be written as
\begin{equation}
\delta ^{(2)} (H+DP)_{u_c} =\frac {1}{2} \int_{-\infty }^{\infty } 
<v,\hat{L}v> d\xi \hspace{.5mm} >0
\end{equation}
where the operator $\hat{L}$ is given by Eq. (32). This means that, if the 
condition in Eq. (37) is fulfilled, then $H(u)+DP(u)$ has a minimum at $u=u_c$.
Inversely, the minimum of $H(u)+DP(u)$ at $u=u_c$ is a sufficient condition 
of compacton stability with respect to small perturbations. Thus from Eq. (30) 
we obtain the condition for the minimum of the perturbed Hamiltonian 
$H_c(\lambda )$ at $\lambda =1$ as 
\begin{equation}
2DP_c(k-1)(k-7) > 64\alpha J_2 (k+1)
\end{equation}
Using Eq. (20), (21) and the exact compacton solutions [Eq. (17)] it can be shown 
that the condition in Eq. (41) is satisfied for arbitrary values of nonlineariy 
parameter $k$ within the range $2\leq k \leq 5$ for which the compacton 
solutions are allowed. This result, that the condition for the 
compacton stability is satisfied for arbitrary values of the nonlinearity 
parameter, is unlike the soliton stability results, where it has been 
shown that the stability condition of the soliton solutions puts a 
restriction on the allowed values of the nonlinearity parameters [5,6,9,10].

We would like to mention here that the stability condition [Eq. (39)] is 
obtained by assuming that at sufficiently small dispersion, 
there is only one eigenstate with negative eigenvalue for the operator 
$\hat{L}$ [Eq. (32)]. Details regarding this conjecture are explained 
in [5]. The validity of this conjecture has been proven from 
numerical experiments for many other systems, such as third and fifth-order 
Korteweg-de Vries equations as well as nonlinear Schr\"odinger equations [5]
. At present we do not have any evidence to show that this conjecture is 
also valid for our operator $\hat{L}$ [Eq. (32)], except for the fact that 
the stability result that follows from using this conjecture also 
agrees with the results 
obtained independently from the Hamiltonian minimum condition  
[Eqs. (40) and (41)]. Numerical experiments along the lines as refered in [5] 
will be required to verify the validity of this conjecture for the systems 
described here.

\section{Conclusions}
To conclude, in this paper we have shown how the nonlinear dispersion term 
interacts with the nonlinear convection term in the fifth-order nonlinear 
dispersive equations to generate exact compacton solutions free from 
exponential tails and many other unusual nonlinear localised solutions like 
peakons, cuspons etc. Using simple scaling relations and the invariance 
property of the equations under stretching group as well as 
spiral group of transformations, we have shown how these higher order 
nonlinear dispersive equations support self similar solutions of various 
patterns. Unlike the third-order nonlinear dispersive 
equations [2], for the fifth-order nonlinear dispersive equations case, 
various solitary wave solutions such as cuspons, peakons, tipons etc 
cannot be shown as a plot in phase diagrams as functions of the 
corresponding potentials.  
However, using an ansatz, we have been able to obtain the 
peakon solutions of the fifth-order nonlinear dispersive equations. 
We have also shown that the fifth- 
order nonlinear dispersive equations when expressed in the form of the 
$K(m,n,p)$ equations [Eq. (2)] have four 
conserved densities ($Q$'s in [Eq. (15)])
same as that for
the third-order nonlinear dispersive $K(m,n)$ equations [Eq. (1)], though the 
corresponding flux densities ($X$'s in [Eq. (15)]) are obviously different. 
Further, even for the arbitrary odd order nonlinear dispersive equations 
we have proved the existence of three conservation laws and provided strong
evidence for the existence of the fourth one.
 However, it 
should be noted that, for the 
case of the fifth-order nonlinear dispersive equations [Eq. (4)] which 
are derivable from a Lagrangian, there are 
only three conserved quantities [3]. We have used
linear stability analysis to examine the stability of the compacton solutions
 for the fifth-order 
nonlinear dispersive equations. The important differences between soliton and 
compacton solutions that come out from the stability analysis of 
the corresponding solutions are that, whereas the soliton solutions are allowed
 for arbitrary values of the nonlinearity parameters, the stability 
condition of the soliton solutions puts restrictions on the range of the 
nonlinearity parameters for which stable soliton solutions are allowed 
. On 
the other hand, the compacton solutions are allowed only 
within a certain range of the nonlinearity parameters  
and all the allowed compacton solutions within this specific range of 
the nonlinearity parameters are stable. This is because, as 
shown above, as in the case of the 
third-order nonlinear dispersive equations [8], the linear stability analysis
of the compacton solutions of the fifth-order 
nonlinear dispersive equations also does not  put any additional constraint 
on the range of the nonlinear parameters.
\newpage
\section{References}
\begin{enumerate}
\item P. Rosenau and J.M. Hyman, Phys. Rev. Lett. {\bf 70}, 564 (1993); 
P. Rosenau, Phys. Rev. Lett. {\bf 73}, 1737 (1994). 
\item P. Rosenau, Phys. Lett. A {\bf 230}, 305 (1997)
\item B. Dey, Phys. Rev. E {\bf 57}, 4733 (1998)
\item P. Rosenau, Phys. Lett A, {\bf 252}, 297 (1999) and references therein.
\item V.I. Karpman, Phys. Lett. A {\bf 210}, 77 (1996) and references therein
\item B. Dey, A. Khare and C.N. Kumar, Phys. Lett. A {\bf 223}, 449 (1996)
\item F. Cooper, J.M. Hyman and A. Khare, e-print Patt-Sol/9704003
\item B. Dey and A. Khare, Phys. Rev. E {\bf 58}, R2741 (1998)
\item E.A. Kuznetsev, Phys. Lett. A {\bf 101}, 314 (1984)
\item V.I. Karpman, Phys. Lett. A {\bf 215}, 254 (1996) and references therein.
\end{enumerate}
\end{document}